# Second- and Third-Harmonic Backscatter through a Bandstop Filter Using Defected Ground Structure

Dan Che, Changjun Liu, *Senior Member, IEEE*, Haoming He, Zhi Hua Ren, and Pengde Wu, *Member, IEEE*

*Abstract*—In this brief, a novel harmonic-backscatter-rectifier (HBR) is introduced for simultaneous rectification and harmonic-based uplink. The proposed (HBR) employs a rectifier to generate both DC power and the harmonic carriers for backscattering communication, and a dual-band reconfigurable band-stop filter using defected ground structure (DGS) to modulate the second and third harmonics with low-power consumption and same input impedance at $f_0$ all the time. As a proof of concept, an HBR prototype operating at 1.85 GHz was designed and experimented. An uplink data rate of 8 Mbps (4 Mbps × 2) was achieved when the HBR was fed with -10 dBm RF power, and the data modulation consumed less than 27.7 pJ/bit. In addition, a passive harmonic tag was implemented with the proposed HBR and a low-power binary sequence generator, which demonstrated a continuous uplink of 12 kbps at -6 dBm RF power.

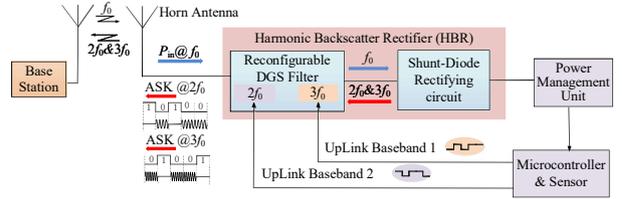

Fig. 1. System illustration of the proposed harmonic backscatter rectifier, where the reradiated harmonics can be modulated in amplitude through a dual-band reconfigurable band-stop filter using DGS.

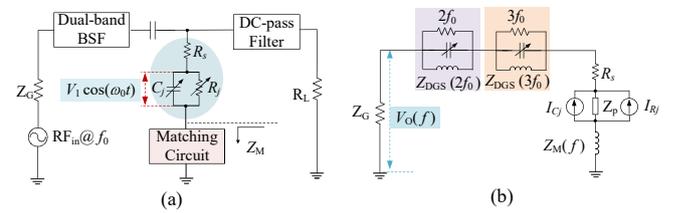

Fig. 2. (a) Block diagram of the proposed HBR. (b) Equivalent circuits describing the proposed HBR at harmonic frequencies.

## I. INTRODUCTION

WITH the boom of the Internet of Things (IoT), an enormous number of wireless sensors has been deployed to collect data in various applications. How to keep such a massive number of sensors active is a great challenge. Traditional batteries have a limited lifespan, and the replacements of batteries would cause enormous maintenance and environmental costs. Thus, it is highly desirable to make battery-free sensor nodes by harvesting power from remote radio frequency (RF) sources and decreasing the power consumption of sensor nodes [1]–[3].

Recently, wireless power transmission (WPT), especially the rectifier has made great progress in high efficiency, wide band, miniaturization and so on [4], [5]. However, traditional radio modules used in IoT sensors comprise power-hungry RF chains that include oscillators, mixers, and digital-to-analog converters, which result in significant power consumption. A particularly promising method for low-power wireless uplink is backscattering communication, which allows IoT sensor nodes to transmit data by reflecting and modulating an incident RF wave instead of generating a new one [6], [7]. Although conventional backscatter communication has been extensively used in RF identification system (RFID) to enable passive tags, self-jamming and multi-path interference are major problems that limit the capabilities of the conventional RFID system and increase the design complexity [8], [9].

Harmonic backscattering communication has gained increasing attention because of its advantages of low cost, high reliability, and low power consumption [10]–[12]. By exploiting the nonlinearity of the rectifier circuit, the excited harmonics are used as uplink carriers, which could reduce self-jamming and multipath interference by separating the uplink communication frequency and the downlink WPT frequency. So, the harmonic based uplink has emerged as a potential solution for long readout range in a harsh environment [13].

Harmonic-based uplink communication primarily utilizes amplitude shift-keying (ASK) modulation techniques, which can be achieved through various methods such as input matching/mismatching [12], RF switches [13], bias of varactor [14], [15], and variation of the DC termination impedance [16]. However, the issues of energy loss due to impedance mismatch at $f_0$ and dedicated harmonic generation have received increasing attention in recent years. To address the issue of DC energy loss caused by impedance mismatch, a bandpass filter (BPF) was used to extract the second harmonic component modulated by the tag baseband signals [17]. Building on this concept, we propose a novel approach in this brief where the reradiated second and third harmonics are modulated simultaneously through a bandstop filter (BSF) utilizing a defected ground structure (DGS), as illustrated in Fig. 1. One of the significant advantages of the proposed method is that it does not need extra circuits for harmonics extraction and duplexing, enabling concurrent energy harvesting and uplink information transmission. Additionally, two uplink channels can be utilized with distinct functionalities, such as one for sensor data transmission and the other for antenna misalignment.

This work was supported by NSFC under Grant 62101366 and Grant 62071316. *(Corresponding authors: Pengde Wu.)*
D. Che, C. Liu, H. He, and P. Wu are all with the School of Electronics and Information Engineering, Sichuan University, Chengdu 610064, China (e-mail: pengdewu2018@gmail.com).
Z.H. Ren is with Zuckerman Institute, Columbia University, NY 10027, New York, USA.

## II. Theoretical Analysis

Fig. 2(a) shows the block diagram of the proposed harmonic backscatter rectifier. The impedance of the matching circuit is denoted by $Z_M$. The equivalent circuit of the diode consists of series resistance $R_s$, junction capacitance $C_j$, and junction resistance $R_j$. At harmonic frequencies, the rectifying diode becomes a $2f_0/3f_0$ energy generator modeled by a Norton equivalent circuit, which is shown in Fig. 2(b). The impedance of two pairs of DGS cells used in the dual-band reconfigurable BSF are denoted by $Z_{\text{DGS}}(2f_0)$ and $Z_{\text{DGS}}(3f_0)$, respectively.

As shown in Fig. 2(b), the backscattered signals on the input side with antenna impedance $Z_G$ are:

$$V_O(2f_0) = \frac{Z_G \times I_{2f_0} \times Z_{p2}}{Z_G + Z_{\text{DGS}}(2f_0) + Z_{p2} + R_s + Z_M(2f_0)} \quad (1)$$

$$V_O(3f_0) = \frac{Z_G \times I_{3f_0} \times Z_{p3}}{Z_G + Z_{\text{DGS}}(3f_0) + Z_{p3} + R_s + Z_M(3f_0)} \quad (2)$$

where $Z_{p2}$ and $Z_{p3}$ are the internal impedance of the current source at $2f_0$ and $3f_0$, respectively. $I_{2f_0}$ and $I_{3f_0}$ are the second and third harmonic currents generated by the junction resistance and the charge stored in the junction capacitance [11].

From (1) and (2), the reradiated $2f_0$ and $3f_0$ power are inversely proportional to $Z_M(f)$. To increase the signal-to-noise ratio (SNR) of the backscattered $2f_0$ and $3f_0$ on the base station side, $Z_M(2f_0)$ and $Z_M(3f_0)$ should have low values to maximize harmonic generation. Considering that the path loss of $3f_0$ is usually larger than that of $2f_0$ with the same antenna gain, $Z_M(3f_0)$ should be lower than $Z_M(2f_0)$. Besides, the imaginary part of the diode must be compensated to achieve impedance matching at $f_0$ [18].

Based on the above discussion, a short-circuited stub with a characteristic impedance of $Z_1$ and an electrical length of $\lambda_g/6$ is used as the diode matching circuit. Its input impedance at different frequencies (DC, $f_0$, $2f_0$, $3f_0$) can be obtained as

$$Z_M(f) = jZ_1\tan\left(\frac{\pi}{3}\frac{f}{f_0}\right) = \begin{cases} 0, & f=0 \\ j\sqrt{3}Z_1, & f=f_0 \\ -j\sqrt{3}Z_1, & f=2f_0 \\ 0, & f=3f_0 \end{cases}. \quad (3)$$

Then (3) shows that the backscattered $3f_0$ power can be maximized because $Z_M(3f_0) = 0$, and the capacitive impedance of the diode can be balanced because of the inductive impedance of short-circuited $\lambda_g/6$ at $f_0$. In addition, the rectifying circuit of the proposed HBR intends to maximize the power of backscattered harmonics for a higher SNR, so its PCE exhibits a drop of ∼5% (additive) when compared to a Class-C rectifier with short-circuit terminations at both the second and third harmonics [19].

On the other hand, the ASK modulation of $2f_0$ and $3f_0$ can be achieved based on the relations in (1) and (2), where the reradiated $2f_0$ or $3f_0$ are also inversely proportional to the impedance of the DGS resonators. When the DGS resonator is in-resonance, $Z_{\text{DGS}}(f) \to \infty$, resulting in much less harmonic power back to the antenna, that is, symbol "0" is sent. When the resonator is off-resonance, $Z_{\text{DGS}}(f) \to 0$; hence, much more harmonic power returns to the antenna, that is, symbol "1" is sent.

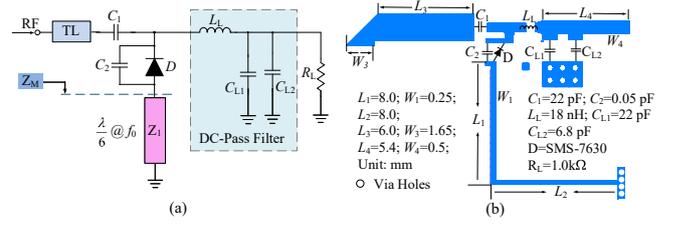

Fig. 3. Shunt-diode rectifying circuit used in HBR: (a) schematic and (b) layout with the design parameters.

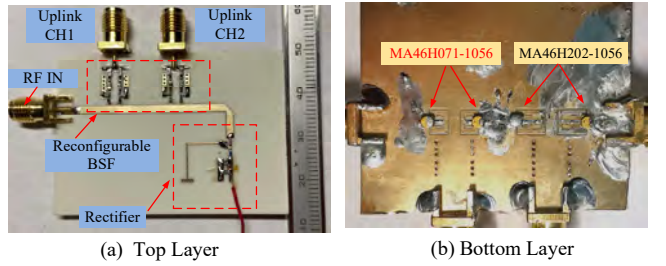

(a) Top Layer      (b) Bottom Layer

Fig. 4. Photograph of the fabricated HBR: (a) top view and (b) bottom view.

## III. Implementation and Experimental Results

### A. Circuit Design

For the experimental validation, an harmonic backscatter rectifier prototype operating at 1.85 GHz was implemented, and an RF laminate RO4350B ($h$=0.762 mm, $\epsilon_r = 3.66$, and $\tan\delta = 0.002$) was used as the substrate. Fig. 4 shows a photograph of the fabricated HBR, in which the rectifier is integrated with the dual-band reconfigurable BSF.

A detailed schematic and layout of the rectifier are shown in Fig. 3, where a small capacitor $C_2$ is employed to reduce the real part of the diode impedance to 50 Ω [20]. The physical dimensions of the DGS resonators used in the reconfigurable BSF are shown in Fig. 5. Two pairs of symmetrical coupled resonators could provide two resonance frequencies: the pair of DGS resonators on the left-hand side is for signal rejection at $3f_0$, whereas the two larger resonators on the right-hand side are for the stop-band at $2f_0$.

The dual-band reconfigurable BSF using the symmetric DGS structure is shown in Fig. 6. To achieve voltage-controlled band positions at $2f_0$ and $3f_0$, as seen in Fig. 6, direct connections between the DGS island and ground plane are cutoff by slots, and then four varactors are mounted between the DGS islands and ground plane [21]. The band position can be tuned as a result of the change in the equivalent capacitance of a DGS resonator.

The frequency response of the dual-band reconfigurable BSF was measured at different bias states. As shown in Fig. 7, the insertion loss at 1.85 GHz was about 0.21 dB, regardless of bias. The initial center frequency of the first stop-band was 3.4 GHz and that of the second stop-band was 4.9 GHz. As indicated in Fig. 7(a), $|S_{21}|$ at $2f_0$ decreased from −2.5 to −18.3 dB when the $2f_0$ DGS cell was biased at 3 V. Similarly, the second resonant frequency shifted from 4.9 to

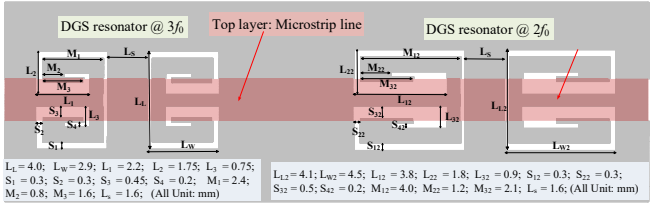

Fig. 5. Physical dimensions of the DGS resonators used in the reconfigurable BSF.

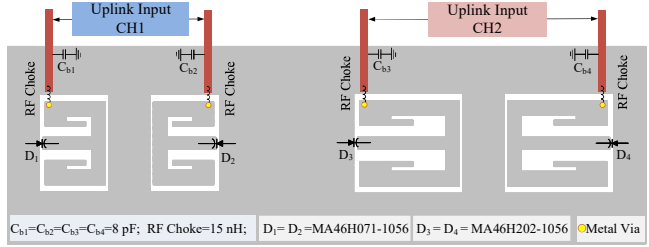

Fig. 6. Bottom view of the reconfigurable BSF with varactors and bias circuits.

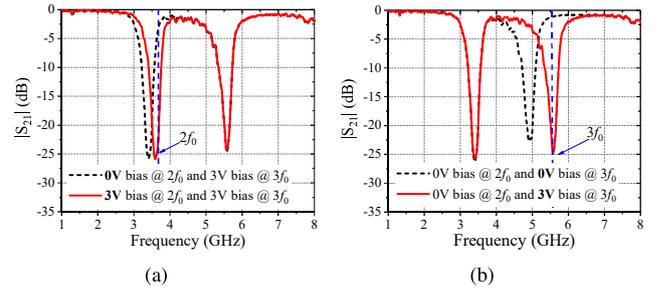

Fig. 7. Resonance frequency shifts as a function of the reversal bias voltage: (a) $2f_0$ resonator and (b) $3f_0$ resonator.

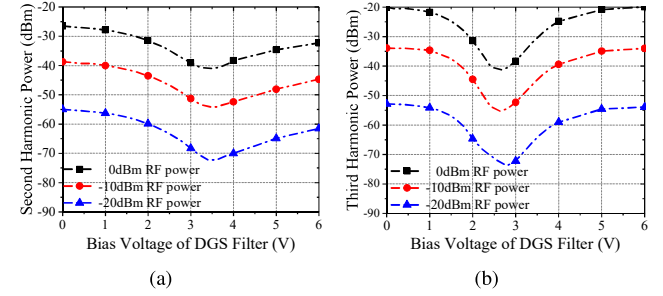

Fig. 8. Measured harmonic power versus bias voltage for (a) $2f_0$ and (b) $3f_0$.

5.55 GHz ($3f_0$) when a bias of 3 V was applied to the $3f_0$ DGS cell, which resulted in $|S_{21}|$ at $3f_0$ decreasing from $-1.1$ to $-23.7$ dB. It was also observed that the controllability of the first and second stop-bands was irrelevant, which indicates that simultaneous amplitude modulation over $2f_0$ and $3f_0$ is possible.

### B. Cabled Experiment

To validate modulation and rectification with the proposed harmonic backscatter rectifier, cabled experiments were conducted. The HBR was fed by an RF signal generator and the modulated second and third harmonics were coupled to a spectrum analyzer with ASK demodulation function (20 dB directional coupler). A dual channel function generator provided the uplink basebands, and square waves with a 50% duty cycle (equivalent to binary sequences) were fed to the bias ports of the reconfigurable BSF. The DC load was $R_L$=1 kΩ in all following experimental tests.

First, static biasing of the reconfigurable BSF was performed to quantify the backscattered harmonic power. The harmonic power versus bias voltage at 0, -10 and -20 dBm are depicted in Fig. 8. As shown, 3.7 GHz power decreased steadily with the bias voltage before hitting the bottom at 3.3 V bias. Similarly, 5.55 GHz power also decreased as the bias voltage increased before reaching the lowest level at 2.8 V bias. Considering particle circuit implementation where the harvested power is stored on a capacitor, to send a "1" symbol, a bias voltage of 2.8 to 3.3 V can be used for moderate modulation depths of 3.7 GHz and 5.55 GHz.

To test the data transmission performance of the proposed HBR, two 0 to 3.0 V square waves were applied to the $2f_0$ and $3f_0$ DGS cell, both square waves had a frequency of 2 MHz. Fig. 9 shows the recovered square wave from the $2f_0$ and $3f_0$ carriers when the HBR was fed by -10 dBm RF power. It is seen that the uplink information of 2 MHz was well recovered by the spectrum analyzer owning a noise floor of about -95 dBm. Thus, a total uplink data rate of 8 Mbps was achieved with uplink at both $2f_0$ and $3f_0$. It should be noted that those waveforms in Fig. 9 represent raw analog data from an envelope detector of the spectrum analyzer.

To verify the effect of harmonic modulation on rectification, the RF-to-DC power conversion efficiency (PCE) of HBR was measured with and without bias. Fig. 10(b) shows the results, where a slight impact is observable in general. The degradation of the PCE at 1.85 GHz was less than 1.23% (additive) at 0 dBm RF power. Additionally, Fig. 10(a) plots the measured output DC voltage with and without biasing. The results show negligible degradation: the output DC voltage dropped from 0.777 to 0.769 V at 0 dBm RF power. In conclusion, the proposed modulation method using a reconfigurable BSF had a slight impact on rectification, facilitating concurrent energy harvesting and uplink information transmission.

Finally, the power consumption of the proposed modulation method was characterized by including the power required for biasing the varactors in four DGS cells. The total work performed to charge the junction capacitance $C_j$ to reverse voltage $U = V_R$ can be obtained from the integral

$$W = \int_0^{W(Q)} dW = \int_{U=0}^{U=V_R} d(U \times Q_j). \quad (4)$$

From (4), the energy consumption required to charge a MA46H202-1056 varactor (in the $2f_0$ DGS cell) from 0 to 3 V was calculated as 20.9 pJ according to its capacitance-voltage variation characteristic [22], while discharging the varactor to 0 V did not consume any energy. As there were two varactors in the $2f_0$ DGS cell, the average energy consumption for sending alternating "0" and "1" was 20.9 pJ. Similarly, the average power consumption for biasing MA46H071-1056 varactors in the $3f_0$ DGS cell was calculated as 6.79 pJ/bit. Considering the use of $2f_0$ and $3f_0$ DGS cells for dual-channel simultaneous data modulation, the total power consumption for modulating $2f_0$ and $3f_0$ was $27.7$ pJ/bit.

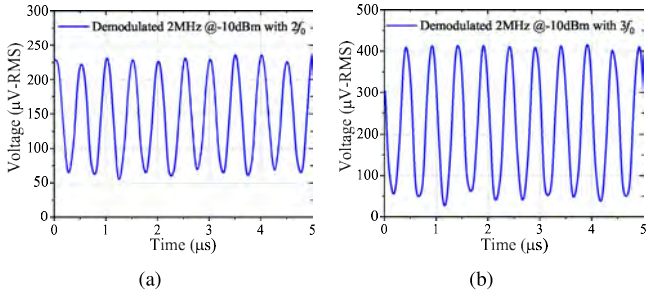

Fig. 9. Demodulated $2\,\text{MHz}$ square-wave from (a) $2f_0$ carrier and (b) $3f_0$ carrier; the spectrum analyzer is set to zero-frequency span mode, the Y-axis is in linear scale and its unit is root-mean-square (RMS) value in volts.

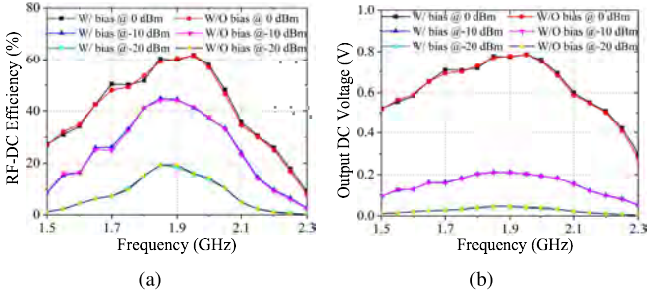

Fig. 10. Measured PCE and output DC voltage versus RF frequency with/without bias voltage, (a) PCE, (b) output DC voltage.

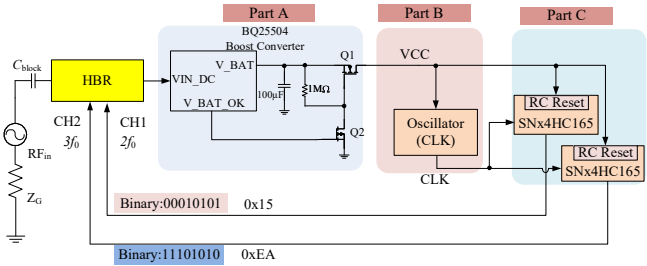

Fig. 11. Experimental measurement setup for the RF-powered uplink with HBR, a power management unit, and a binary sequence generator.

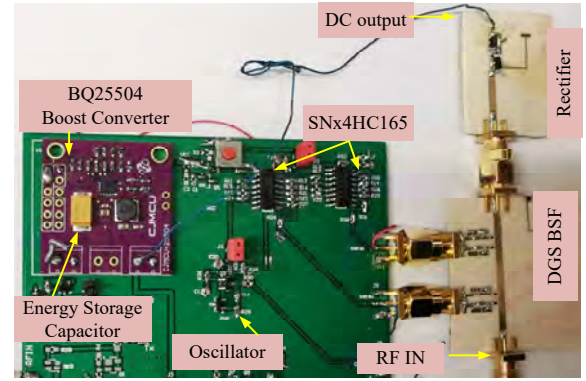

Fig. 12. Implementation of an RF-powered harmonic tag with the proposed HBR, a power management unit and a binary sequence generator.

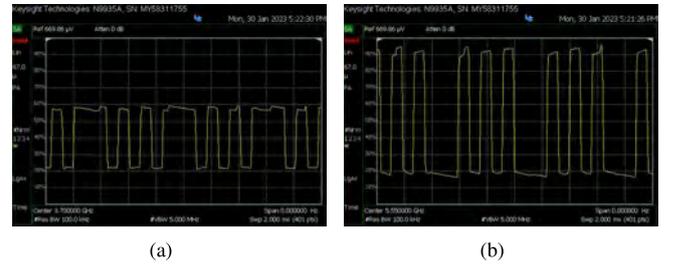

Fig. 13. Demodulated data from (a) $2f_0$ carrier and (b) $3f_0$ carrier with an RF-powered HBR, the data are ones' complement of 0x15 and 0xEA.

*C. RF-powered Uplink*

In order to better demonstrate the potential of the proposed HBR for low-power IoT applications, a RF-powered harmonic tag was designed, as shown in Fig. 11. It consists of the proposed HBR, a power management unit (Part A, bq25504), an RC oscillator (Part B), and parallel-in/ parallel-out shift registers (Part C, SNx4HC165×2). Part B and Part C will work together as a binary sequence generator.

Fig. 12 is a photograph of the tag prototype. As shown, the output DC power of the rectifier is directed to the boost converter "bq25504" located at the upper-left corner. The RC oscillator, which consists of only two transistors, is in the middle part of the tag. Serial outputs from the shift registers are sent directly to the dual-band reconfigurable BSF.

In the proposed design, the boost converter was designed to have lower and upper voltage thresholds of $2.7$ V and $3.3$ V, respectively. When the voltage across the $100\,\mu\text{F}$ capacitor ($V_{\text{BAT}}$) reaches $3.3$ V during RF energy harvesting, PMOS will be turned on and the energy will flow from the capacitor to the binary sequence generator. In the meanwhile, the power-on reset circuits enable the parallel-in mode of the shift register, then preset hex digits (0x15 and 0xEA, representing ID address and collected sensor data) will be loaded into the shift registers asynchronously on power up, and they will also be clocked out in an endless loop since those binary numbers are also reloaded to the serial input (SER) of SNx4HC165. The experimental results show that the power consumption of the binary sequence generator is $63.5\,\mu\text{W}$ with a $3.0$ V power supply and $12$ kbps data rate.

The HBR was fed by -6 dBm RF power in the demonstration of RF-powered uplink. It took $2.28$ s to charge the $100\,\mu\text{F}$ capacitor from $2.62$ to $3.28$ V with a total energy of $194.7\,\mu\text{J}$ transferred from RF signal to capacitor, so the average output power of the boost converter is about $85.4\,\mu\text{W}$ and the RF-DC efficiency of the proposed HBR with a "bq 25504" is $34.2\%$. Considering that the power consumption of the binary sequence generator is $63.5\,\mu\text{W}$, the harvested DC power is sufficient for the continuous modulation over $2f_0$ and $3f_0$. Fig. 13 shows the recovered square wave from the $2f_0$ and $3f_0$ carriers when the binary sequence generator was powered by the harvested energy under -6 dBm RF input power. It is shown that the data rate is about $12$ kbps, and the demodulated data are ones' complement of the generated binary numbers.

A comparison between this study and other low-power uplinks based on rectifier nonlinearity is listed in Table I. The results show that, compared with existing solutions for amplitude modulation over the backscattered harmonics, the proposed reconfigurable DGS filter enables simultaneous rectification and uplink with only one rectifier. Moreover, the capability of two channels of uplink with carrier of $2f_0$ and $3f_0$, which is more promising for clock synchronization in high-speed wireless communication.

TABLE I
COMPARISON WITH PREVIOUS WORKS ON LOW-POWER UPLINKS

| Ref. | [14] 2019 | [16] 2022 | [23] 2018 | This Work |
|---|---|---|---|---|
| Objective | Harmonic RFID | Harmonic transponder | WPT+ uplink | WPT+ uplink |
| Uplink carrier | 2nd harmonic | 2nd harmonic | IM3* | 2nd & 3rd harmonic |
| WPT Fre. | 434MHz | 1.04 GHz | 4.94 & 5.0 GHz | 1.85 GHz |
| Uplink Fre. | 868MHz | 2.08 GHz | 5.06 GHz | 3.7& 5.55GHz |
| Modulation type | Amplitude | Amplitude | Amplitude | Amplitude |
| Modulation Method | Bias of varactor | dc impedance of doubler | Bias of varactor | DGS filter |
| Harmonic generation | NLTL* | Frequency doubler | Rectifier | Rectifier only |
| Power/Signal Routing | Power split | N.A. | N.A. | Collision-free |
| Interruption to WPT | Yes | No dc output | Yes | No |
| Power Cons. of modulation | N.A. | 221pJ/bit | N.A. | 27.7pJ/bit |
| Data Rate | 65.4 kbps | 2 Mbps | 100 kbps | 8 Mbps |
| *: IM3: third-order intermodulation; NLTL: nonlinear transmission lines ||||||

## IV. CONCLUSION

In this brief, a novel harmonic backscatter rectifier was proposed to achieve uncompromised rectification and low-power uplink communication at the same time. The backscattered second and third harmonics were modulated by a reconfigurable BSF using DGS in amplitude simultaneously, and an uplink data rate of 8 Mbps (4 Mbps×2) was demonstrated in a cabled experiment, with 27.7 pJ/bit energy efficiency. The measured PCEs with and without biasing only demonstrated a drop of 0.8% (additive) at -10 dBm RF power, which proves that the modulation has a negligible effect on rectification. In addition, an RF-powered harmonic tag was implemented with the proposed HBR, a power management unit and a binary sequence generator, and a continuous uplink of 12 kbps was demonstrated by using the harvested DC power from -6 dBm RF signal.

In contrast to the published modulation methods used for amplitude modulation over backscattered harmonics, the proposed dual-band DGS filter enables a continuous power flow to the wireless tag with a combination of backscatter communications. However, due to the fabrication of four DGS resonators etched on the ground plane, the proposed DGS filter requires a larger circuit area. This issue may be addressed by integrating the rectifier with the DGS filter using a custom CMOS based IC fabrication technology.